\documentclass[12pt]{article}
\usepackage{pic04}
\usepackage{hyperref}
\usepackage{url}
\usepackage{graphicx}

\RequirePackage[usenames]{pstcol}

\definecolor{gray12}{rgb}{1.0, 1.0, 0.8}
\definecolor{brightblue}{rgb}{0.0,0.5,1.0}

\begin{document}

\title{\bf The D$\emptyset$ Silicon Track Trigger}
\author{
Lars Sonnenschein        \\
{\em Boston University} \\
on behalf of the D$\emptyset$ collaboration \\
{\em }}
\maketitle

\baselineskip=14.5pt
\begin{abstract}
The Level-2 Silicon Track Trigger preprocessor (L2 STT) of the D$\emptyset$ detector in
Run II is described. It performs a precise reconstruction of charged particle tracks
in the Central Fiber Tracker (CFT) and the Silicon Microstrip Tracker (SMT). Events with
displaced tracks originating from the decay of long living particles such as $B$ hadrons
are triggered on. The presence of $b$ quarks contained in such hadrons is relevant for
$B$ physics and crucial as signature of top quark and Higgs boson decays.
\end{abstract}

\baselineskip=17pt

\section{Introduction}
In Run II the Fermilab Tevatron $\bar{p}p$ collider operates at a
center-of-mass energy of $\sqrt{s}=1.96\,\mbox{TeV}$. 
%In comparison to Run I
%the collision rate ($=7.6\,\mbox{MHz}$) increased by a factor of ten.
To keep the rejection of background events while maintaining high efficiency
for physics processes of interest, the D$\emptyset$ trigger had to be upgraded
and accommodated to the decreased beam crossing time of $132\,\mbox{ns}$ which
requires minimized dead-time between collisions. The D$\emptyset$ Silicon Track Trigger
(STT) \cite{d0c98} is the newest addition to this upgrade.

\section{D$\emptyset$ tracking and trigger} 
The D$\emptyset$ trigger consists of three levels. 
The first level is hardware based and compares data with preprogrammed patterns.
%On the hardware based first level
%energy clusters in the calorimeter, preprogrammed hit patterns in the Central Fiber Tracker
%(CFT) and in the muon chambers are located. 
%Spatial correlations between detector elements are computed. 
The STT belongs to the second level trigger
%, consisting of
%preprocessors and a global Level 2 processor which makes trigger decisions 
%based on information received from the preprocessors.
%To each major D$\emptyset$ detector component corresponds a Level 2 preprocessor. The STT
and constitutes a preprocessor for the Silicon Microstrip Tracker (SMT). The output of 
Level 2 preprocessors is sent to the third trigger level which applies
sophisticated reconstruction algorithms to the data.
% on a computer farm.
To maintain a dead-time below five percent the mean decision time for the Level 2 trigger
has to be kept below $100\,\mu\mbox{s}$ subdivided in two halfs for the preprocessors 
and the global L2 decision.

The STT is fed with information from silicon strip detectors arranged in four concentric layers 
in six cylindrical barrels around the beam axis (fig. \ref{schematic}, left). 
All detectors have axial $50\,\mu\mbox{m}$ 
pitch strips (parallel to the beam line). 
%Double-sided detectors have in addition 
%60-150$\,\mu\mbox{m}$ pitch stereo strips ($2^{\circ}$ or 
%$90^{\circ}$ rotated to axial strips). 
The STT subdivides the inputs from the barrel 
detectors into six independent azimuthal $60^{\circ}$ sectors.
% since the loss of
%high transverse momentum tracks hitting detectors belonging to neighbored sectors 
%is negligible. 
In addition to the SMT information the STT receives 
up to 46 tracks per sector each event from the Level 1 Central Track Trigger (CTT)
which makes trigger decisions on tracks from the Central Fiber Tracker (CFT).

The STT builds clusters from the SMT raw hits and defines 
%a $\pm 2\,\mbox{mm}$ road 
$\pm 2\,\mbox{mm}$ wide roads around
the Level 1 CTT tracks (fig. \ref{schematic}, right). 
%Axial clusters are used for STT track fitting. 
%Stereo clusters are passed on to Level 3. 
The CFT hits of the inner- and outermost layers
are used together with SMT clusters of at least three out of four layers for the STT
track fitting. The determined track fit parameters are sent to the Level 2 CTT/STT trigger
and to Level 3.

\section{STT hardware design}  
\begin{figure}[t]
  \centerline{\hbox{ \hspace{1.8cm}
      \includegraphics*[bb=100 100 600 800, angle=270, width=9.0cm]{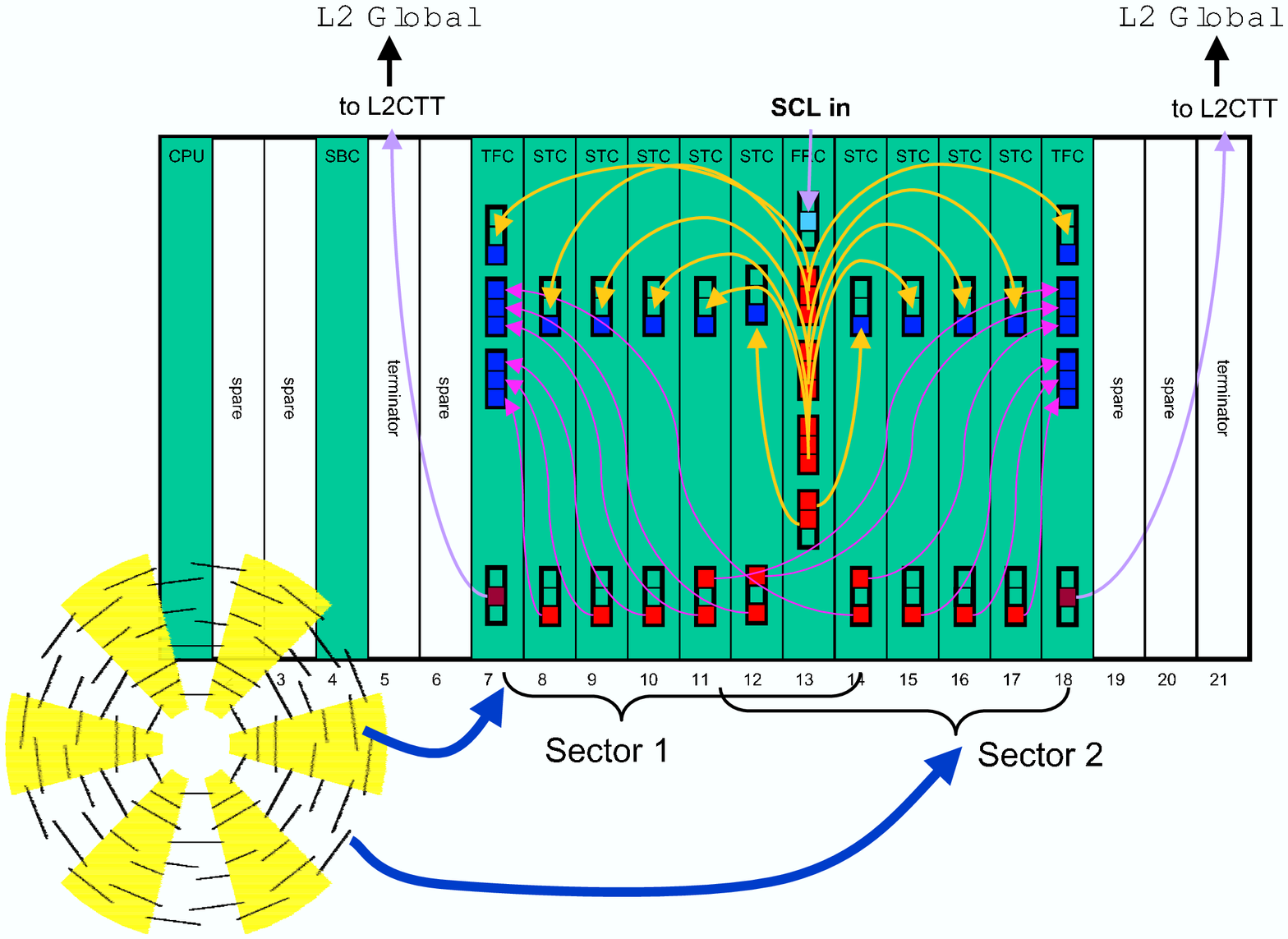} 
      \hspace{-0.3cm}
      \raisebox{-5.5cm}{\includegraphics[angle=90, width=10cm]{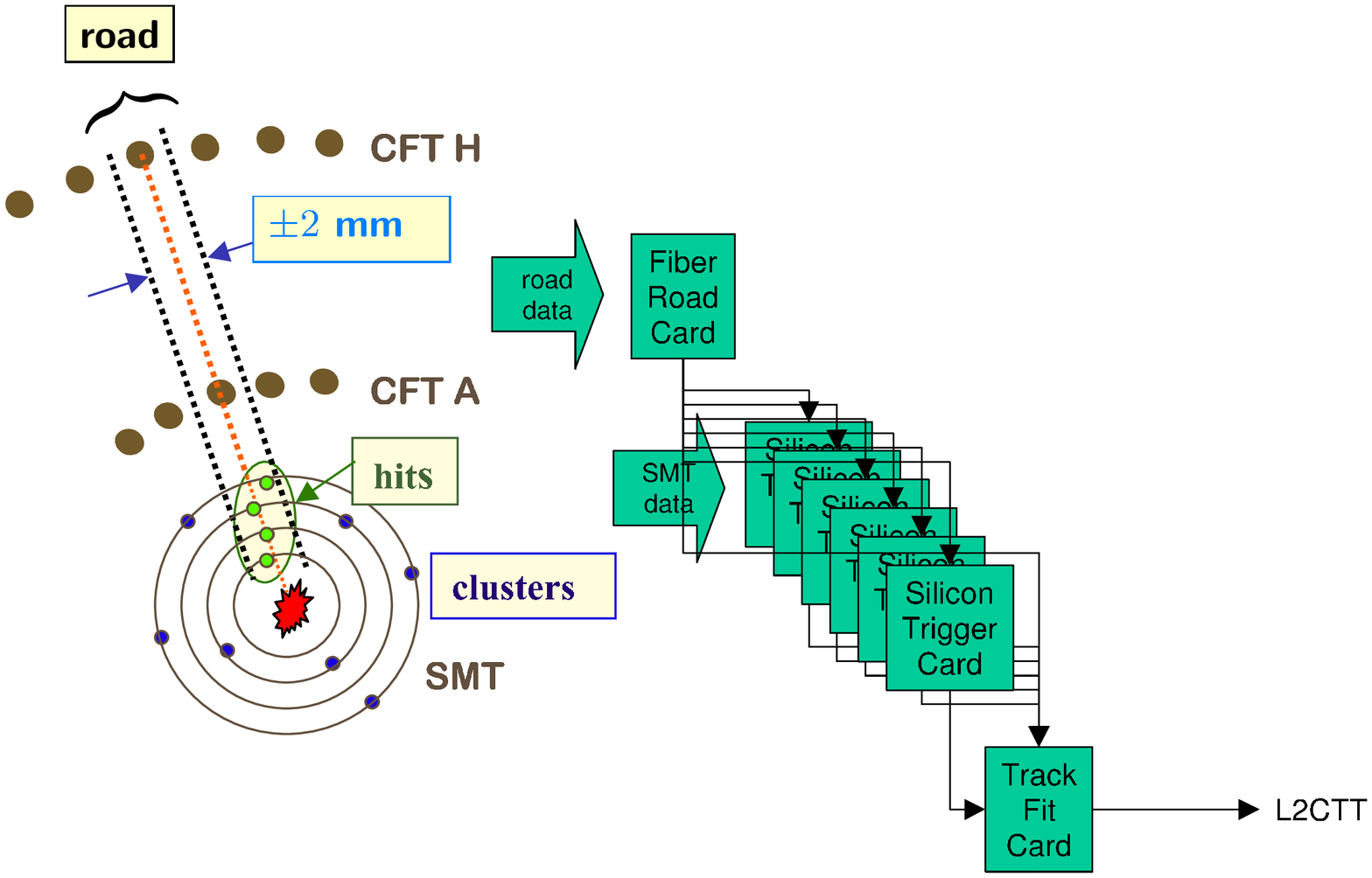} }
    }
  } \vspace*{-0.6cm}
  \caption{\it
    STT crate with SMT barrel detectors (left) 
    and STT road definition with SMT hit selection and subsequent data flow (right).
    \label{schematic} }
\end{figure}

The STT consists of six identical crates, each covering two neighbored $30^{\circ}$ azimuthal 
SMT sectors (fig. \ref{schematic}, left).
A crate caries an input output controller (IOC) to operate the crate, 
a single board computer (SBC) for Level 3 data submission 
and twelve custom-designed VME boards with programmable processors and daughter boards.
%which are one Fiber Road Card, nine Silicon Trigger Cards,
%two Track Fit Cards, 18 Buffer Controllers (one BC each VME board) 
%and 31 Low Voltage Digital Signal (LVDS) Link Receiver (LRB) and Transmitter Boards (LTB).
%While data is communicated serial between different boards via LTB's and LRB's
%control information uses dedicated backplane signals.

%\subsection{Motherboard}
The STT motherboard is a 9U$\times$400$\,$mm VME64x-compatible 
%\cite{vme97}
card containing three 33$\,$MHz PCI 
%\cite{pci98} 
busses 
% - for simultaneous data transmission - 
to communicate
between the logic daughter board 
(either a Fiber Road Card (FRC), 
a Silicon Trigger Card (STC)
or a Track Fit Card (TFC)),
serial link boards %(LTB, LRB)
and the data buffer controller (BC) board.
 
%It has a VIPA \cite{vip99} standard P0, P1 and P2
%connectors while the J3/P3 connector is non-standard to accommodate the backplane of
%the existing SMT readout system.

%The FRC, STC and TFC daughter boards with their unique logic are 
%connected to the other daughter boards and the backplane via
%three PCI busses which in turn are connected via two independent PCI-to-PCI bridges
%for simultaneous data transmission.
%The VME bus at the backplane is used for readout to Level 3, initialization and monitoring. 
%A VME-to-PCI bus bridge on the motherboard allows the IOC CPU to access daughter boards.

%\subsection{Fiber Road Card}
The FRC consists of four functional elements which are
a trigger receiver which communicates with the trigger framework bidirectional, 
a road receiver which receives tracks from the L1 CTT trigger, 
a trigger/road data formatter which combines the road and trigger framework information
and a buffer manager which controls the buffering and readout to Level 3.
% physically implemented in three separate 
%Altera FLEX10K Field Programmable Gate Arrays (FPGA). 

\begin{figure}[b]
  \centerline{\hbox{ \hspace{1.9cm}
      \includegraphics*[angle=0, width=6.6cm]{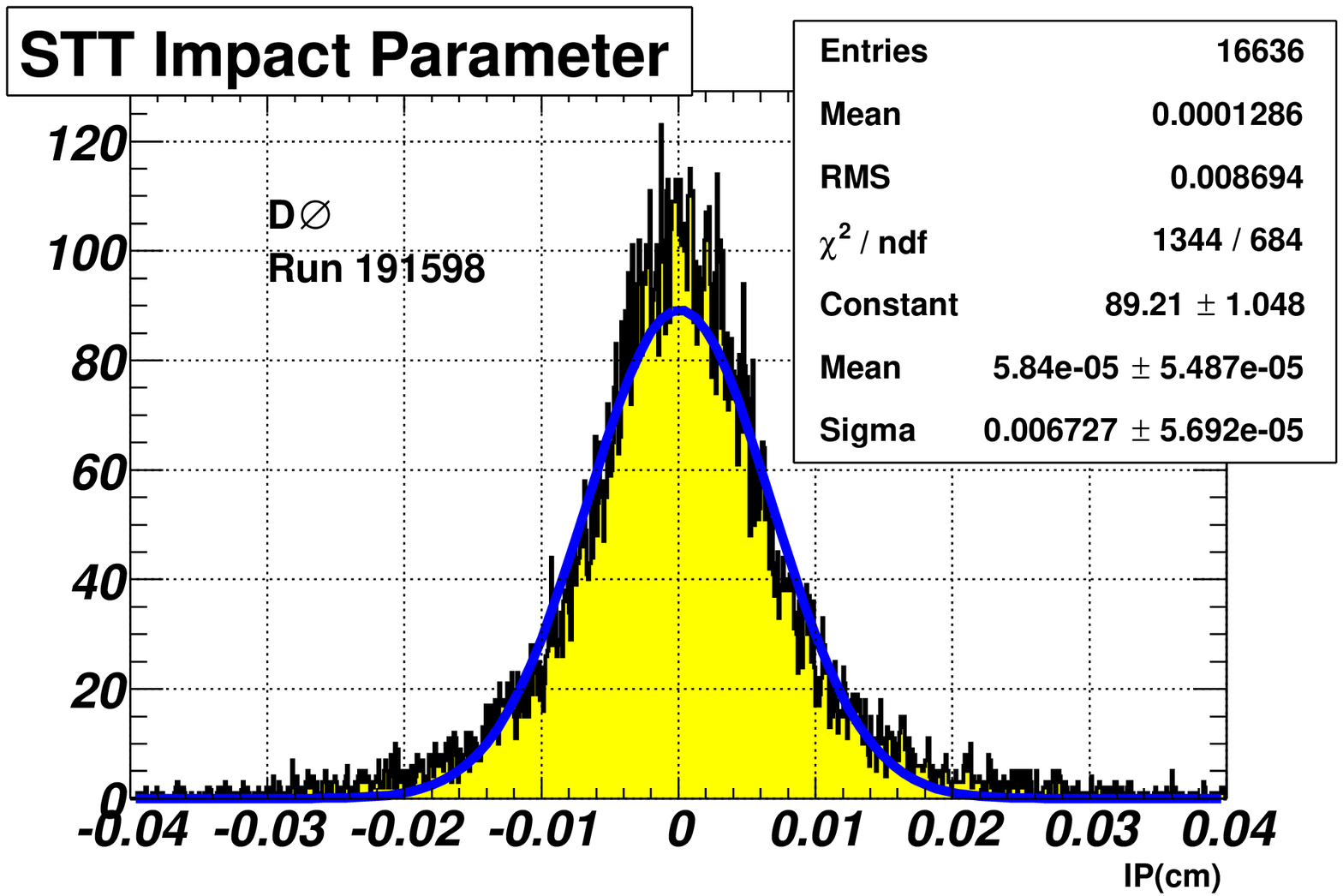} 
      \hspace{-0.2cm}
      \includegraphics[angle=0, width=6.6cm]{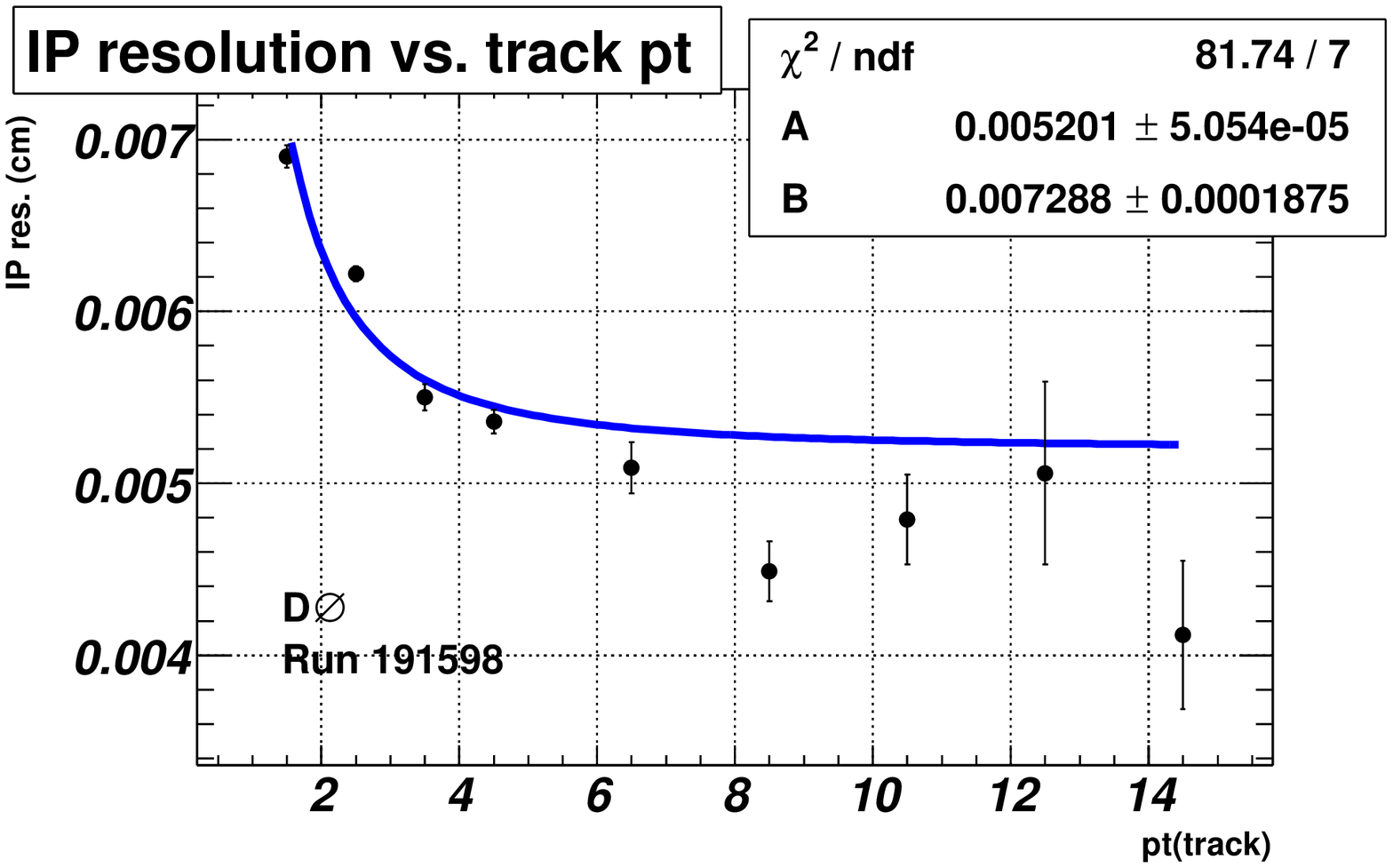}
      \hspace{-0.1cm}
      \raisebox{4.1cm}{\includegraphics*[bb=290 370 600 800, angle=270, width=6.2cm]{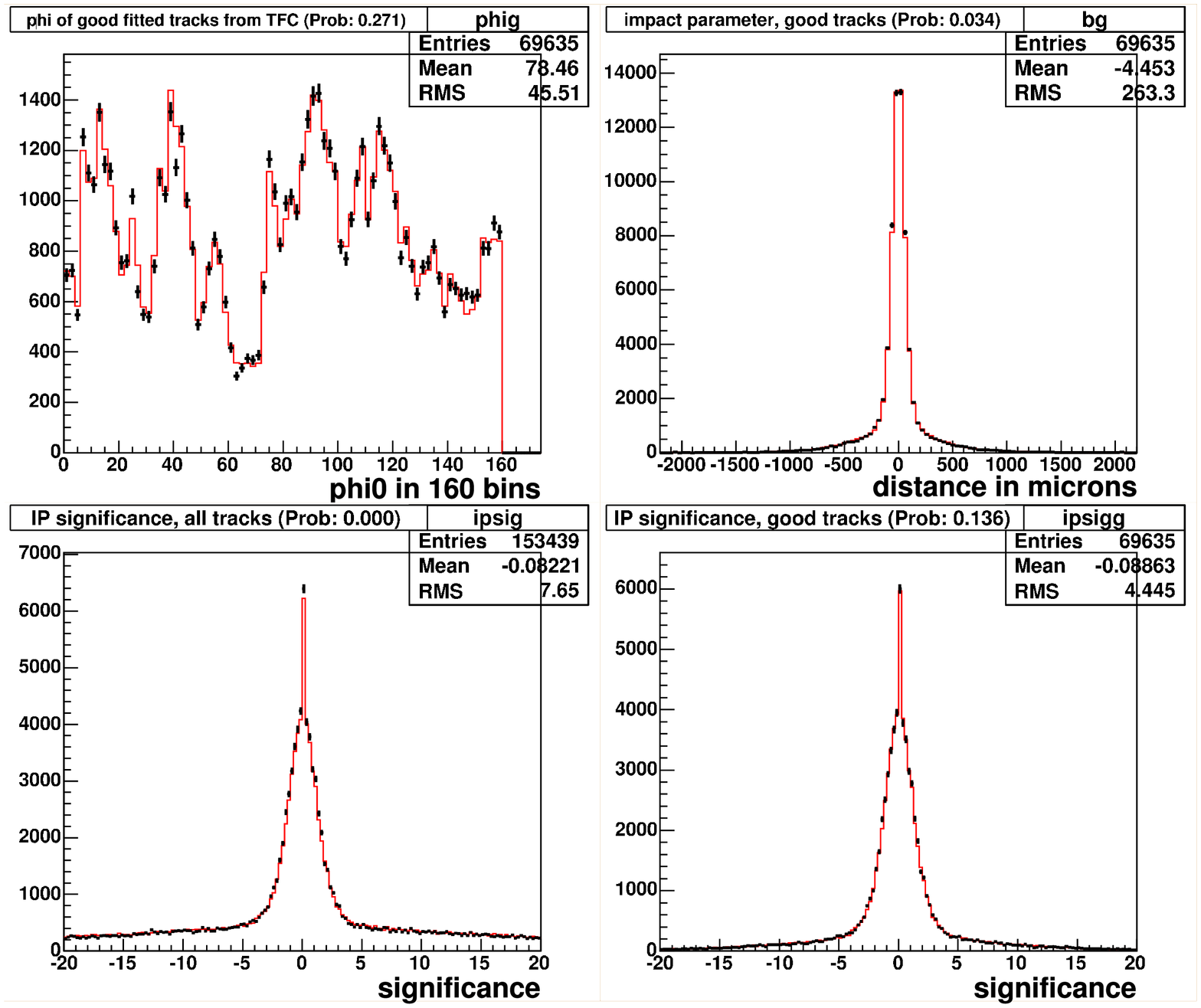} }
    }
  } \vspace*{-0.4cm}
  \caption{\it
    Impact parameter (IP) resolution (left), 
    IP resolution versus track transverse momentum in GeV$/c^2$ (center) and
    IP significance (right) - measured with data (preliminary).
    \label{dataplots} }
\end{figure}

The STC receives raw hits from the SMT barrel detectors. 
Bad strips are masked and pedestal/gain corrections are applied via Look-Up-Tables (LUT).
Adjacent SMT hits are clustered. Clusters of axial strip detectors are matched to roads around
CFT tracks via LUT's and sent to the Track Fit Card (TFC).

The TFC performs the final cluster filtering (closest cluster to road center) 
and two dimensional linearized track fitting of the form 
\begin{equation}
  \phi(r) = b/r + \kappa r + \phi_0  
\end{equation}
where $r$ is the radial distance from the beam spot position to the cluster, 
$\phi$ its azimuth, $b$ the impact parameter,
$\kappa$ the track curvature and $\phi_0$ its direction at the point of closest approach.
The fit is performed with help of precomputed matrix elements stored in LUT's
exploiting two CFT hits in addition to the SMT clusters.

\section{Performance}

The impact parameter of the STT integrated over all track transverse momenta
has been measured to be $67\,\mu\mbox{m}$ (fig. \ref{dataplots}, left, preliminary) 
including the beam spot size of about
$35\,\mu\mbox{m}$ and the $15\,\mu\mbox{m}$ spatial resolution of the axial 
SMT ladder detectors, obtained offline.

The measured impact parameter resolution of the STT as a function of track transverse momentum
(fig. \ref{dataplots}, center, preliminary) increases towards lower track transverse momenta
due to multiple Coulomb scattering. The fit function 
\begin{equation}
  \mbox{IP}=\sqrt{A^2+(B/{p_{\perp}})^2} 
\end{equation}
has been used.

Fig. \ref{dataplots} (right) 
shows a monitoring plot with the impact parameter significance
distribution of the STT. The red curve is a reference histogram, the black dots are data.
In the present test trigger a cut on events with at least one track 
with impact parameter significance above $+3$ has been implemented.

\section{Conclusions}
The STT measures impact parameters of tracks and their significance 
which allows to select large samples of events containing long living particles 
such as $B$ hadrons in the presence of enormous background through triggering on 
displaced tracks. Presently the STT operates
as a test trigger during data taking and will be fully commissioned with the next 
revision (V14) of the D$\emptyset$ trigger list.

\section{Acknowledgments}
Many thanks to the staff members at Fermilab, collaborating institutions
and the Altera and Xilinx Corporations for contributions.
Major project funding has been granted from the Department Of Energy (DOE)
and the National Science Foundation (NSF) under the Major Research Instrumentation
(MRI) Program Award No. 9977659,9/1/1999.

\end{document}